# A stochastic user-operator assignment game for microtransit service evaluation: A case study of Kussbus in Luxembourg

[a]Tai-Yu Ma, [b]Joseph Y.J. Chow, [a]Sylvain Klein, [b]Ziyi Ma

*Abstract—* This paper proposes a stochastic variant of the stable matching model from Rasulkhani and Chow [1] which allows microtransit operators to evaluate their operation policy and resource allocations. The proposed model takes into account the stochastic nature of users' travel utility perception, resulting in a probabilistic stable operation cost allocation outcome to design ticket price and ridership forecasting. We applied the model for the operation policy evaluation of a microtransit service in Luxembourg and its border area. The methodology for the model parameters estimation and calibration is developed. The results provide useful insights for the operator and the government to improve the ridership of the service.

## I. INTRODUCTION

Microtransit service has received increasing interests as an alternative to provide users (travelers) with flexible and convenient mobility services. This kind of service is characterized by its route flexibility to meet users' needs in terms of pickup and drop-off locations, and additional service such as wifi connections allowing passengers to work while traveling. Generally, a microtransit service operator deploys a fleet of vehicles to provide either door-to-door or a combined service to transit stations based on users' need. It has been tested in several cities in North America and Europe as a good solution for solving the first and last-mile transportation problems [2]-[3]. Integrating microtransit as a part of Mobility-as-a-Service (MaaS) further enhances seamless multimodal transportation service to reduce personal car use. However, operating microtransit services usually requires higher operation cost compared to classical mass transportation which relies mainly on the subsidy from the government. Operators like Via and MaaS Global along with failed microtransit services like Kutsuplus [4], Bridj [5], and Chariot [6] show the importance of jointly taking into account the interactions of users' preference and operators' operating policy to evaluate the ridership and revenue on the service network.

Most studies address the MOD/microtransit services planning from the supply-side perspective, focusing on vehicle routing and dispatching [7]–[12], pricing [13], vehicle rebalancing [14]. Existing methodology for modeling user-transportation network interactions is usually based on dynamic traffic assignment models (DTA) or dynamic systems simulation via day-to-day adjustment [15-18]. These methodologies rely on modeling user's travel choice behavior to achieve market equilibrium. However, in systems involving microtransit or mobility-on-demand service, it is difficult to draw insights on the impact of cost allocation mechanism on ridership using traditional traffic assignment methods which requires using agent-based simulation model to derive traffic flow on the network.

Different from the existing DTA approach, Rasulkhani and Chow [1] proposed a new user-route assignment approach based on the assignment games [19], [20]. This approach considers user-operator/-route assignment as a matching problem given users' travel preference and operators' cost allocation mechanism. Under this modeling approach, users pay a ticket price for using the service and receive a net utility. Operators gain a net revenue by reducing its operating cost. The assignment game problem is designed such that users and operators have sufficient incentives (non-zero profit on each side) to participate. The recent development of the stable matching model further considers user-operator assignment on a multimodal transportation market with multiple operators [21]. However, there is still no realistic applications on microtransit service evaluation.

In this study, we propose a stochastic variant of the stable matching model of Rasulkhani and Chow [1] for microtransit service evaluation. First, we present the proposed stochastic user-operator assignment model to match users and a set of service lines with capacity constraints. The model considers users' stochastic travel utility perception resulting in a probabilistic stable operation cost allocation problem to set up ticket prices and forecast ridership. Second, we develop the methodology to estimate the model parameters and calibrate them based on the empirical microtransit service Kussbus (https://kussbus.lu/) data shared by the microtransit operator UFT (Utopian Future Technologies S.A.) in Luxembourg. Third, we apply the proposed approach to an empirical study of Kussbus in Luxembourg and its border areas. The impact of different pricing policies, route cost reduction and vehicle capacities on ridership and operator's profit is evaluated which provided useful insights for the operator to improve their service planning.

## II. METHODOLOGY

Notation

| | |
|---|---|
| $s \in S$ | a user or a set of homogeneous users |
| $r \in R$ | Index of routes, defined as a sequence of stops visited by a shuttle (bus) |
| $U_{sr}$ | Utility/payoff gained for user $s$ for matching route $r$. |

* The work was supported by the Luxembourg National Research Fund (INTER/MOBILITY/17/11588252), NSF grant CMMI-1634973, and C2SMART University Transportation Center.
[a]T. Y. Ma and S. Klein are with Luxembourg Institute of Socio-Economic Research (LISER), 11 Porte des Sciences, L-4366 Esch-sur-Alzette, Luxembourg (Tel:+352 585855308; fax: +352 585855100; e-mail: tai-yu.ma@liser.lu; sylvain.klein@liser.lu).
[b]J. Y. J Chow and Z. Ma are with C2SMART University Transportation Center, New York University, 15 MetroTech Center, Brooklyn, NY11201, USA (e-mail: joseph.chow@nyu.edu; ziyi.ma@nyu.edu).

| | |
|---|---|
| $C_r$ | Operation cost of route r |
| $t_{sr}$ | Generalized travel cost for user s taking route r to connect origin to destination. It could be measured as the weighted sum of walking, waiting and in-vehicle riding time and ticket price. |
| $a_{sr}$ | Payoff gained resulting from $(s,r)$ match, defined as $a_{sr} = \max(0, U_{sr} - t_{sr})$ |
| $u_s$ | Payoff gained for user s |
| $v_r$ | Profit gained for route r |
| $u_{sr}$ | Payoff gained for user s from matching route r |
| $v_{sr}$ | Profit gained for the operator from the user-route matching $(s,r)$ |
| $c_{sr}$ | Operation cost for user s to use route r |
| $p_{sr}$ | Ticket price for user s to use route r |

We consider a set of users to be assigned to a set of routes provided by a microtransit operator. A route is defined as a sequence of stops visited by a bus/shuttle. The user-route assignment problem is modeled as a many-to-one assignment game in which each route can be matched with multiple users and each user can be matched with at most one route. Moreover, vehicle capacity needs to be satisfied. An illustrative example is shown in Figure 1 in which we consider one user and two routes. A user's generalized travel cost is considered as a door-to-door travel cost, measured as a weighted sum of walking time, waiting time, in-vehicle travel time, and ticket price paid to the operator. We call users as buyers and operators as sellers. The objective of the assignment game is to find a seller-buyer matching/assignment such that a total generalized payoff is maximized. The output of the stable matching model is the user-route flow and resulting ticket prices and the operating cost allocation of routes. The model allows testing the impact of different operation strategies and pricing on the ridership and operator's revenue. The reader is referred to [1] for a more detailed description of the model properties.

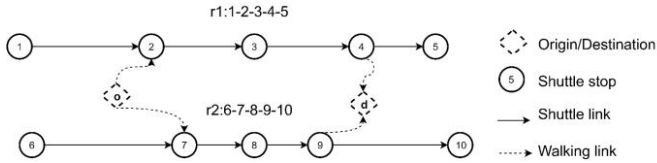

Figure 1. Example of one user and two routes (r1 and r2).

The stable matching model is formulated as follows. First, an optimal user-route assignment game problem (P1) is formulated to find user-route flows that maximize total generated payoff on the operator's service network.

**P1: User-route assignment model**

$$\max \sum_{s \in S} \sum_{r \in R} a_{sr} x_{sr} \quad (1)$$

s.t.

$$\sum_{r \in R} x_{sr} \leq q_s, \forall s \in S \backslash \{k\} \quad (2)$$

$$\sum_{s \in S \backslash \{k\}} \delta_{asr} x_{sr} \leq u_r, \forall a \in A_r, r \in R \quad (3)$$

$$\sum_{s \in S \backslash \{k\}} x_{sr} \leq M(1 - x_{kr}), \forall r \in R \quad (4)$$

$$x_{sr} \in \{0, \mathbb{Z}_+\}, \forall s \in S, \forall r \in R \quad (5)$$

The objective function (1) is to maximize total payoff gains form the assignment. The payoff gained for user-route pair $(s,r)$ is $a_{sr} = \max(0, U_{sr} - t_{sr})$, where $U_{sr}$ is the utility gain of $(s,r)$, $t_{sr}$ is the generalized travel cost of $(s,r)$. Equation (2) states the assignment of users on routes is constrained by its demand. Equation (3) states route capacity constraint $u_r$. $\delta_{asr}$ is an indicator being 1 if arc $a$ is used by user $s$ for route $r$ and 0 otherwise. $k$ is a dummy user of not matching with any route with a utility of 0. Equation (4) ensures that a route is only matched when its total payoff exceeds a threshold cost. $M$ is a big positive number. Equation (5) ensures that the decision variable $x_{sr}$ is a non-negative integer.

We extend the model of [1] by assuming the utility $U_{sr}$ as an independent random variable composed of a deterministic part $V_{sr}$ and an unobserved part $\varepsilon_{sr}$ as

$$U_{sr} = V_{sr} + \varepsilon_{sr} \quad (6)$$

where $V_{sr}$ is the mean utility gain and $\varepsilon_{sr}$ is a random utility term that follows a Normal distribution with mean 0 and standard deviation $\sigma$.

Given the assignment result of P1, a stable cost allocation problem (P2) is formulated under a desired objective in which operators and users have no incentive to switch. The output of P2 is the profile of net payoffs for users and routes of operators.

**P2: User-operator stable cost sharing model (stable sharing model)**

$$\max Z \quad (7)$$

s.t.

$$\sum_{s \in G(r,x)} u_s + v_r \quad (8)$$
$$\geq \sum_{s \in G(r,x)} a_{sr} - C_r, \forall G(r,x) \text{ and } r \in R$$

$$\sum_{s \in S(r,x)} u_s + v_r = \sum_{s \in S(r,x)} a_{sr} - C_r, \forall r \in R^* \quad (9)$$

$$v_r = 0, \forall r \in R \backslash R^* \quad (10)$$

$$u_s = 0, \quad if\ s \in \bar{S} = \{s | \sum_{r \in R} x_{sr} = 0\} \quad (11)$$

$$u_s \geq 0, v_r \geq 0, \forall r \in R^* \quad (12)$$

Equation (7) is a user-defined cost allocation mechanism to be maximized. A common setting is to use the buyer-optimal and seller-optimal objectives to obtain the vertices for the full range of stable outcomes. For the buyer-optimal case, the total utility gain of users is maximized, i.e. $Z = \sum_{s \in S} u_s$ to maximize; for the seller-optimal case, the total profit gain of operators is maximized, i.e. $Z = \sum_{r \in R} v_r$. Equation (8) is the stable condition for which no user would have a better payoff other than the current assignment. $G(r, x)$ is the group of users that can be feasibly assigned on route $r$ given the solutions $x$ of P1. Note that when operating costs can freely transfer between routes, Eq. (8) can be relaxed. We assume this constraint holds for the Kussbus case study. Equation (9-11) are the feasibility conditions where $R^*$ is the subset of routes with at least one matched user. $S(r, x)$ is the set of users matching route $r$, given an optimal assignment solution $x$ obtained by P1. $u_s$ and $v_r$ are non-negative continuous decision variables. The solution of P2, denoted as $\{(u, v); x\}$, is a cost allocation outcome given $x$.

By introducing (6) in (8), (8) becomes:

$$\sum_{s \in G(r,x)} u_s + v_r + C_r - \sum_{s \in G(r,x)} V_{sr} \geq \sum_{s \in G(r,x)} \varepsilon_{sr}, \quad (13)$$
$$\forall G(r, x), r \in R$$

Given $\varepsilon_{sr}$ follows a Normal distribution, its summation is also a Normal distribution. We introduce the concept of $\alpha$-stability to measure the reliability for which the stability constraint is satisfied (e.g. $\alpha = 0.05$ implies being 95% sure of matching satisfying the constraint). Equation (13) can then be expressed deterministically as a chance constraint (14).

$$\Phi \left[ \frac{(\sum_{s \in G(r,x)} u_s + v_r + C_r + \sum_{s \in G(r,x)} V_{sr})}{|G(r,x)|} \right] \quad (14)$$
$$\geq 1 - \alpha, \quad \forall G(r, x), r \in R$$

where $\Phi(z) = \Pr(Z \leq z)$ is the cumulative density function of Z.

By [22], the nonlinear constraint (14) can be transformed to a linear inequality as

$$\sum_{s \in G(r,x)} u_s + v_r + C_r - \sum_{s \in G(r,x)} V_{sr} \geq Z_{1-\alpha} \sigma' \quad (15)$$

with new deviation $\sigma' = \sqrt{|G(r,x)|\sigma^2}$. The extension is a generalization of the deterministic model from [1] with $\alpha = 0.50$. As the P2 problem is a linear programming problem, we can use the simplex algorithm or interior-point algorithm to obtain solutions efficiently.

Given the $\alpha$-stability outcome $\{(u, v); x\}_\alpha$, we can determinate ticket prices as:

$$p_{sr} = v_{sr} + c_{sr}, \forall r \in R, \forall s \in S \setminus \{k\} \quad (16)$$

with $\sum_{s \in S(r,x)} c_{sr} = C_r$ and $\sum_{s \in S(r,x)} v_{sr} = v_r, \forall r \in R$

Given the cost allocation output obtained from P2, ticket prices can be determined using either the equal-share or cost-based share policy. Note that in P1, the ticket price is left out from $a_{sr}$ to determine optimal user-route flows. Ticket pricing is then determined from the outcome of P2. Operators can evaluate the ridership and profit as user-route matched flows by integrating ticket prices as a part of user's generalized travel cost in P1. Consequently, the resulting ridership and operator profit depend on the integrated price schemes.

### III. DATA

Kussbus Smart shuttle service (https://kussbus.lu/en/how-it-works.html) is a first microtransit service operating in Luxembourg and its border area. The service was provided by the Utopian Future Technologies S.A. (UFT) from April 2018 to March 2019. Like most microtransit systems, users use dedicated Smartphone applications to book a ride in advance with desired origin, destination and pickup time as input. Service routes are flexible to meet maximum access distance constraint. Routes are generated in a way that users need to walk from/to the origin/destination to/from shuttle stops given a pre-defined threshold (i.e. around one kilometer). The service started operating between the Arlon region in Belgium and the Kirchberg district of Luxembourg City on 04/25/2018 and a second line started on 09/24/2018 between Thionville region (France) and Kirchberg district. Both service areas are highly congested on road networks due to high car use during morning and afternoon peak hours.

The empirical ride data was provided by the operator for the period from 4/25/2018 to 10/10/2018. A total of 3258 trips (rides) were collected. Each ride contains the following information: booking date and time, pickup time and drop-off time, pickup and drop-off locations, pickup and drop-off stops, walking distance between stops and origins/destinations, origin-destination pairs of users, and fare. Any abnormal trips (e.g. trip duration less or equal to 5 minutes) were removed. As a result, a total of 3010 trips were used for this study.

The characteristics of Kussbus service is summarized as follows.

- Service areas: two service areas: a.) Arlon region (Belgium) ↔ Kirchberg district (Luxembourg City), and b.) Thionville region (France) ↔ Kirchberg district.

- Operating hours: From 05:30 to 09:30 and from 16:00 to 19:00 from Monday to Friday.

- Vehicle capacity: vehicle capacity differs from 7-seater, 16-seater, and 19-seater.

- Booking and ticket price: users need to book a ride by the dedicated Smartphone application. First 6 trips are free, and then the unit ticket price is around 5 euros per trip.

- Vehicle routing policy: vehicle routes are scheduled based on pre-booked customer requests on previous days. Late-requests could be accepted under certain operational constraints.

More detail about the operation policy and characteristics of Kussbus service can be found at: https://uft.lu/en/news/references/kussbus.

Figures 2 show the operating routes from Arlon to Luxembourg City. We can observe that most users are picked up at Arlon and transported to Luxembourg City in the morning. There are 429 possible routes observed from Arlon to Kirchberg and 449 in the reverse direction. From Thionville to Kirchberg there are 52 routes with 50 routes in the reverse direction. Figure 3 shows the cumulative probability distribution (CDF) of in-vehicle travel time. 60% of trips have a duration between 40 to 60 minutes. Figure 4 shows the weekly ridership during the study period. Note that the fall of ridership observed for the last two weeks was due the holidays of All Saints' Day.

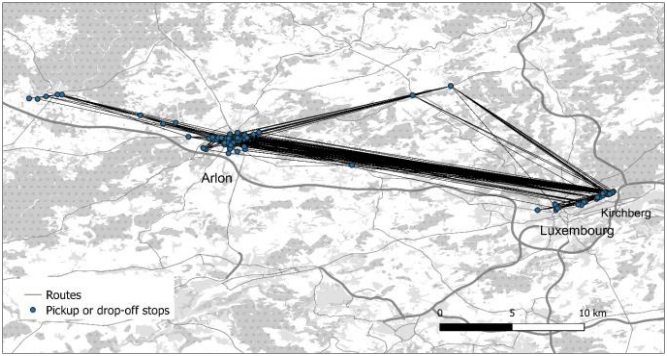

Figure 2. Kussbus operating routes from Arlon to Luxembourg City.

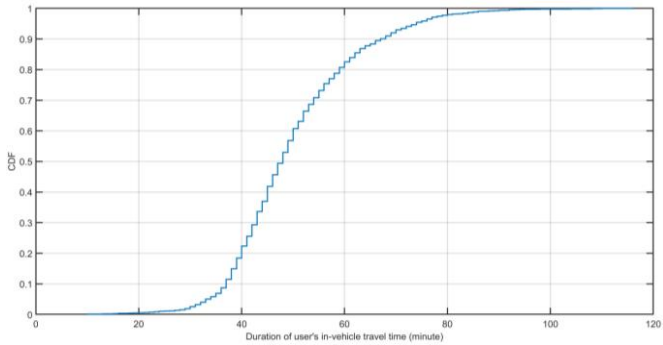

Figure 3. The cumulative probability distribution (CDF) of in-vehicle riding time.

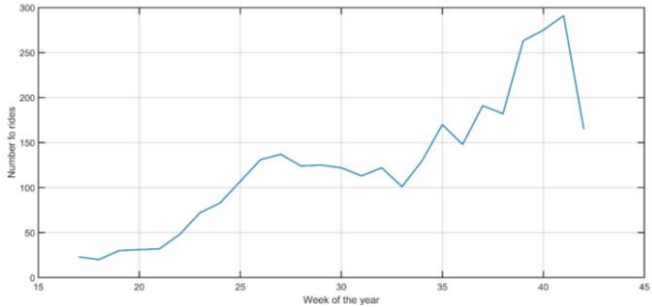

Figure 4. Number of rides per week during 04/2018-10/2018.

## IV. STABLE MATCHING APPLICATION

### A. Case study setting

We apply the proposed model for a realistic microtransit operation evaluation based on the ride data of Kussbus. The data contains 235 morning or afternoon commuting periods. The routes are generated beforehand. The characteristics of Kussbus service are shown in Table 1. The average length of routes is 46.5 km. Three types of shuttles, i.e. 7-, 16- and 19-seater; were used. The route operation cost is estimated as proportional to its travel distance. The average operation cost per kilometer traveled is provided by the operator by taking into account the type of vehicles, fuel cost, driver cost and extra costs. Route travel time is estimated using Google Maps API during morning and afternoon peak-hour to consider congested traffic conditions.

The value of in-vehicle travel time (VOT) is estimated by using a mobility survey conducted in October-November 2012 for the EU officials and temporary employees working in the European institutions at the Kirchberg district of Luxembourg. The sample contains information on 370 valid individuals' commuting practice and their socio-demographic attributes. These individuals live in France, Belgium and Germany border areas of Luxembourg that matches perfectly this study.

Based on our previous study [23-26], we specify two econometric models, i.e. the mixed logit model and multinomial probit model for commuting mode choice behavior modeling. As the mixed logit model specification does not allow us to obtain the convergence result, we retain the result of the multinomial probit model for the VOT estimation. The estimated VOT is 24.67 euros/hour which is consistent with the existing study for the VOT estimation of Luxembourg [27].

TABLE I. KUSSBUS SERVICE CHARACTERISTICS AND PARAMETERS SETTINGS.

| Attribute | Value | Attribute | Value |
| --- | --- | --- | --- |
| Number of trips | 3010 | Average travel distance of users | 43.0 km |
| Value of in-vehicle time (VOT) (euro)[a] | 24.21 | User's maximum waiting time at stop | 10 minutes |
| Walking speed | 5km/hr | Capacity of vehicles | 7, 16 and 19 passenger seats |
| Average route distance | 46.5 km | Average route cost | 61.0 euros |

Remark: a. based on the estimation using a mobility survey for the EU officials and temporary employees working in the European institutions in Luxembourg.

### B. Calibrations of utility and α

We calibrate the utility $U_{sr}$ in (1) based on the Kussbus ride data. The data is divided into a first 80% training dataset and a test dataset with the rest 20%. $U_{sr}$ is considered as $U_{sr} = U_s^0 + t_{sr} + \varepsilon_{sr}$, where $t_{sr}$ is a door-to-door generalized travel cost, estimated as a weighted summation of user's access time to shuttle stops, in-vehicle travel time and waiting time. The value of waiting time and walking time is set as 1.5 and 2 times of VOT [27]. $U_s^0$ is a constant utility term to be calibrated.

We vary $U_s^0$ from 0 to 100 and solve the P1 problem. The calibration aims to find a $U_s^0$ which best matches the observed user-route flow in the data. Figure 5 shows $U_s^0 \geq 45$ euros fits observed user-route matches with 79.03% corrected prediction rate on the training data. For the test dataset, it is 65.45%. We retain $U_s^0 = 45$ euros for this case study. Finally, the mean and standard deviation of $U_{sr}$ is 73.39 and 3.57 respectively for Belgium-side rides. For the French-side rides, these numbers become 72.96 and 8.75, respectively.

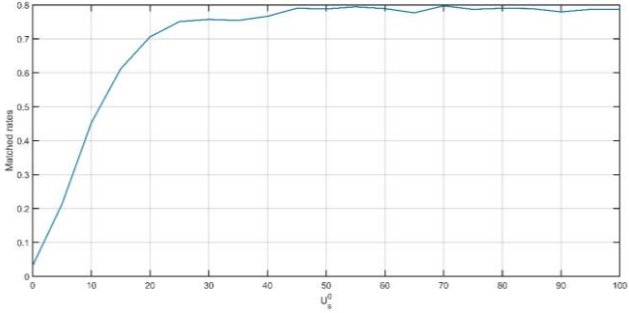

Figure 5. The calibration of constant part $U_s^0$ of trip utility.

## C. Result

We calibrate the reliability parameter $\alpha$ to match observed user-route flow on the data. The buyer-optimal policy is used in P2 to maximize the ridership. The values of $\alpha$ is varied from the set of (0.05, 0.1, 0.2, 0.3, 0.4, 0.5). As shown in Figure 6, we find $\alpha = 0.2$ has best fit with 63.45% corrected prediction rate for the training dataset and 54.43% for the test dataset, respectively. Table II gives the detail user-route assignment result based on Kussbus case study. We use $\alpha = 0.2$ and the test dataset to solve the P2 problem to obtain the ticket prices based on the buyer-optimal and operator-optimal pricing policies. Figure 7 shows the CDF of the ticket prices under different pricing schemes. We find Kussbus's current tariff is quite low which gave 6 free rides to users and then charge around 5 euros per ride. For the buyer-optimal policy, the 50-percentile of the ticket prices is 10.98 euros. For the operator-optimal policy, it becomes 49.91 euros, which is still much cheaper compared to the current taxi fare in Luxembourg.

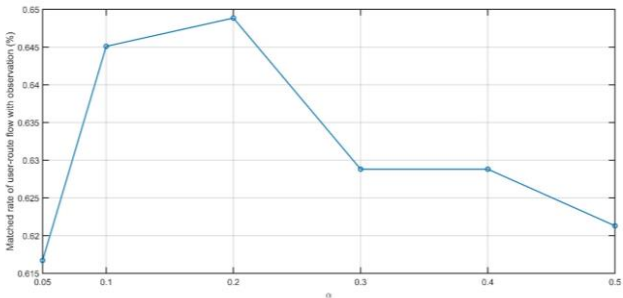

Figure 6. Corrected prediction rates of observed user-route matches for the training dataset over different $\alpha$.

Finally, Table III shows the operator's ridership, revenue, operation cost and profit under these pricing schemes. We find that Kussbus' operated policy would accumulate a financial loss of -4135 euros for 465 matched users. When setting the ticket prices on the buyer-optimal policy, there is 426 users should match with the routes with a positive profit of 187 euros. However, when setting the ticket prices on the operator-optimal policy, there are only 6 users who match with the routes due to high ticket prices. For taxi fare, there are no matches due to its very high tariff with respect to users' trip distance.

TABLE II. USER-ROUTE ASSIGNMENT RESULT OF KUSSBUS RIDES FOR 235 PERIODS.

| Data | Number of ride requests (users) | Number of user-route assignment | User-route assignment rate | Number of rides matched with observations | Matched rate (observation v.s. prediction) |
|---|---|---|---|---|---|
| Training dataset | 2395 | 1829 | 0.7638 | 1520 | 0.6345 |
| Test dataset | 615 | 431 | 0.7011 | 335 | 0.5443 |

Remark: The reported result is the average of 5 runs. The average computational time for the training data set is 84.7 seconds and 29.0 seconds.

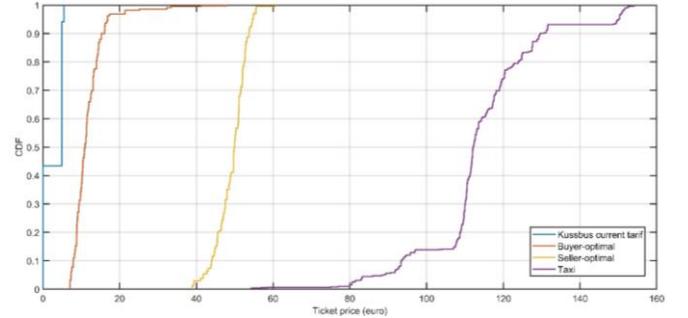

Figure 7. The cumulative probability distribution of user's ticket prices under different pricing policy for the test dataset.

TABLE III. REVENUE, OPERATING COST AND PROFIT OF DIFFERENT PRICING POLICIES FOR THE TEST DATASET.

| Policy | Ridership | Revenue | Operating cost | Net profit |
|---|---|---|---|---|
| Kussbus's tariff | 465 (75.5%) | 1266 | 5401 | -4135 |
| Buyer-optimal ticket price | 426 (69.2%) | 4831 | 4644 | 187 |
| Seller-optimal ticket price | 6 (0.98%) | 231 | 103 | 128 |
| Taxi | 0 | 0 | 0 | 0 |

Remark: Measured in euros. The reported result is the average of 5 runs.

## D. Sensitivity analysis

We conduct a sensitivity analysis to evaluate the impact of route cost reduction and vehicle capacity changes scenarios. Two pricing policies based on the buyer-optimal and seller-optimal setting are evaluated using the test dataset and α=0.2. Table IV shows the ticket prices for different scenarios. For the route cost reduction scenario, under the buyer-optimal policy, we find when reducing from -10% to -50% of the route cost, the ticket price would reduce from -6.9% to -37.9%. However, under the seller-optimal policy, the ticket price would keep stable with less than 1% variation. For the vehicle capacity scenario, we find using a fleet of 7-seater shuttles would reduce the ticket prices up to -9.5% based on the buyer-optimal policy. When increasing the vehicle capacity to 16 and 19 seats, the resulting ticket prices would increase considerably to +56% and +75.1% due to a higher cost for operating these types of vehicles. For the seller-optimal policy, we find a lower ticket price of around -9.4% compared to the current mixed fleet.

TABLE IV. TICKET PRICE VARIATION BASED ON DIFFERENT SCENARIOS.

| Route cost Reduction | BO | | SO | |
|---|---|---|---|---|
| | Euro | ±% | Euro | ±% |
| 0% | 11.6 | | 49.1 | |
| -10% | 10.8 | -6.9 | 49.4 | 0.6 |
| -20% | 10 | -13.8 | 49.4 | 0.6 |
| -30% | 9 | -22.4 | 49.3 | 0.4 |
| -40% | 8.2 | -29.3 | 49.1 | 0.0 |
| -50% | 7.2 | -37.9 | 48.9 | -0.4 |
| Veh. capacity changes | BO | | SO | |
| | Euro | ±% | Euro | ±% |
| Mixed fleet | 11.6 | | 49.1 | |
| 7-seater | 10.5 | -9.5 | 44.5 | -9.4 |
| 16-seater | 18.1 | 56.0 | 44.6 | -9.2 |
| 19-seater | 20.2 | 74.1 | 44.4 | -9.6 |

Remark: BO: Buyer-optimal pricing; SO: Seller-optimal pricing. The result is based on the average of 5 runs.

Table V shows the impact of these two scenarios on the operator's ridership and profit based on the buyer-optimal policy. We find that a route cost reduction of 50% would increase the ridership up to +10%. However, it is not beneficial for the operator. Figure 8 shows the impact of the route reduction scenario on the ridership and profit under buyer-optimal and seller-optimal policies.

For the vehicle capacity scenario under the buyer-optimal policy, using a fleet of 7-seater vehicles would increase the ridership of 3.5%. While for the 16- and 19-seater scenarios, it would reduce the ridership up to -5.6% and -16.2%, respectively, due to the higher ticket prices (see Table IV). The impact on the operator's profit is negative with -103.5% and 106.4%, respectively.

We conclude that if the government were to intervene, we recommend they subsidize Kussbus to improve their route operating costs. Moreover, we find the current mixed fleet strategies are more beneficial compared to using a fleet of vehicles of the same capacity. Our sensitivity analysis shows how the proposed stable matching model can be applied to evaluate different service designs. The operator can apply this methodology to set up ticket prices by considering the price ranges from buyer-optimal and seller-optimal policies.

TABLE V. INFLUENCE OF DIFFERENT SCENARIOS ON THE RIDERSHIP AND PROFIT.

| Route cost Reduction | Ridership | | Profit | |
|---|---|---|---|---|
| | # | ±% | Euro | ±% |
| 0% | 431 | | 202 | |
| -10% | 428 | -0.7 | 199 | -1.5 |
| -20% | 436 | 1.2 | 129 | -36.1 |
| -30% | 460 | 6.7 | 113 | -44.1 |
| -40% | 467 | 8.4 | 148 | -26.7 |
| -50% | 474 | 10.0 | 12 | -94.1 |
| Veh. capacity changes | Ridership | | Profit | |
| | # | ±% | Euro | ±% |
| Mixed fleet | 431 | | 202 | |
| 7-seater | 446 | 3.5 | 0 | -100.0 |
| 16-seater | 407 | -5.6 | -7 | -103.5 |
| 19-seater | 361 | -16.2 | -13 | -106.4 |

Remark: based on the buyer-optimal policy. The result is based on the average of 5 runs.

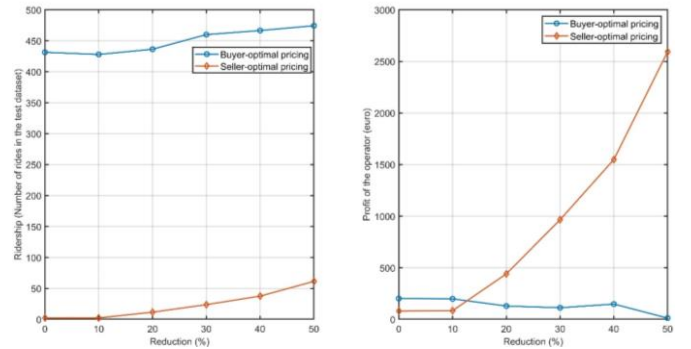

Figure 8. Influence of different route cost reduction scenarios on the ridership and the profit of the operator.

V. CONCLUSION

This study proposes a stochastic user-operator assignment game model for the microtransit operation policy evaluation. The proposed methodology generalizes the previous work of [1] by considering the utility function $U_{sr}$ as a stochastic variable, which converts the deterministic stable matching model to a stochastic one. By introducing the concept of $\alpha$-reliability and formulate the cost allocation problem with chance constraints, the proposed model allows us to better capture the heterogeneity of travel preference for its realistic application for microtransit operation policy evaluation. We conduct the first empirical applications of the model based on the historical ride data from Kussbus microtransit service in Luxembourg and its border areas. The data is shared by the industry collaborator Utopian Future Technologies S.A. covering 3010 trips made between April to October 2018.

We calibrated the model by 80% of the data and validated the model by the rest of 20% of the data. We found Kussbus current ticket prices are not sustainable. By increasing the ticket price to the buyer-optimal policy, it would reduce ridership from the current 465 trips to 426 trips and changing the net profit from -4135 euros to 187 euros for 615 ride requests. We conduct a sensitivity analysis with respect to the operation cost reduction and vehicle capacity changes scenarios. The result shows reducing the operation cost of 50% would increase the ridership of 10% while resulting in further financial loss. We find that the government can intervene by offering to subsidize Kussbus to improve their routing algorithms and reduce operating costs while requiring operation under a buyer-optimal policy. For the vehicle capacity scenario, we find under the buyer-optimal policy, the current mixed fleet strategy would be more beneficial for the operator compared to using a homogeneous fleet of vehicles.

This study shows the potential of the proposed model for on-demand mobility service evaluation. Future extensions could consider the evaluation of shared mobility platforms (see Chapter 3.5 in [28]) and microtransit service in a Mobility-as-a-Service market [21]. Furthermore, considering dynamic cost allocations (e.g. [29]) could be another avenue for future research.

ACKNOWLEDGMENT

We thank the Utopian Future Technologies S.A. (UFT) for providing Kussbus riding data for this research. Particular thanks to Ms. Kimberly Clement for her technical support for the empirical data analysis and to Saeid Rasulkhani for providing the computer program from the previous study and helpful comments.